\documentstyle[12pt,epsfig]{article}

\input epsf

\newcommand{\be}{\begin{equation}}
\newcommand{\ee}{\end{equation}}
\newcommand{\ba}{\begin{array}}
\newcommand{\ea}{\end{array}}
\newcommand{\baa}{\begin{eqnarray}}
\newcommand{\btab}{\begin{tabular}}
\newcommand{\etab}{\end{tabular}}
\newcommand{\eaa}{\end{eqnarray}}
\newcommand{\ci}[1]{\cite{#1}}

\newcommand{\lab}[1]{\label{#1}}

\newcommand{\edd}{\end{document}}

\newcommand{\alf}{\ifmmode\alpha \else$\alpha \ $\fi}
\newcommand{\bt}{\ifmmode\beta \else$\beta \ $\fi}
\newcommand{\gm}{\ifmmode\gamma \else$\gamma \ $\fi}
\newcommand{\Dl}{\ifmmode\Delta \else$\Delta \ $\fi}
\newcommand{\eps}{\ifmmode\varepsilon \else$\varepsilon \ $\fi}
\newcommand{\dl}{\ifmmode\delta \else$\delta \ $\fi}
\newcommand{\et}{\ifmmode\eta \else$\eta \ $\fi}
\newcommand{\vphi}{\ifmmode\varphi \else$\varphi \ $\fi}
\newcommand{\om}{\ifmmode\omega \else$\omega \ $\fi}
\newcommand{\pl}{\ifmmode\partial \else$\partial \ $\fi}
\newcommand{\ps}{\ifmmode\psi \else$\psi \ $\fi}
\newcommand{\sg}{\ifmmode\sigma \else$\sigma \ $\fi}
\newcommand{\phf}{\ifmmode\varphi^4 \else$\varphi^4 \ $\fi}
\newcommand{\Lam}{\ifmmode\Lambda \else$\Lambda$\fi}
\newcommand{\frc}[2]{{#1/#2}}

\newcommand{\ppp}[1]{%
        \setbox0=\hbox{#1}%
        \kern-.02em\copy0\kern-\wd0
        \kern+.04em\copy0\kern-\wd0
        \kern-.02em\raise.0217em\box0}

\newcommand{\lsim}{%less than or approx. symbol
 \mathrel{\setbox0=\hbox{$<$}\raise0.6ex\copy0\kern-\wd0
 \lower0.65ex\hbox{$\sim$}}}

\newcommand{\gsim}{%less than or approx. symbol
 \mathrel{\setbox0=\hbox{$>$}\raise0.6ex\copy0\kern-\wd0
 \lower0.65ex\hbox{$\sim$}}}

\newcommand{\etal}{{\em et al.}}

\newcommand{\PRD}[3]{Phys.\ Rev.\ D {\bf {#1}}, {#2} ({#3})}

\newcommand{\NPA}[3]{Nucl.\ Phys.\ A {\bf {#1}}, {#2} ({#3})}
\newcommand{\NPB}[3]{Nucl.\ Phys.\ B {\bf {#1}}, {#2} ({#3})}
\newcommand{\PLB}[3]{Phys.\ Lett.\ B {\bf {#1}}, {#2} ({#3})}

\newcommand{\JPG}[3]{J. Phys.\ G {\bf {#1}}, {#2} ({#3})}
\newcommand{\EJP}[3]{Eur. J. Phys.\ C {\bf {#1}}, {#2} ({#3})}

\begin{document}
%
%\renewcommand{\thefootnote}{\fnsymbol{footnote}}

%%%%%%%%%%%%%%%%%%%%%% title page %%%%%%%%%%%%%%%%%%%%%%%%%%%%%%
\begin{titlepage}
\renewcommand{\thefootnote}{\fnsymbol{footnote}}
\makebox[2cm]{}\\[-1in]
\begin{flushright}
\begin{tabular}{l}
TUM/T39-99-7\\
%hep-ph/99????
\end{tabular}
\end{flushright}
\vskip0.4cm
\begin{center}
  {\Large\bf
On scale dependence of QCD string operators}
\footnote{work supported in part by BMBF}

\vspace{2cm}

N. Kivel\footnote{Alexander von Humboldt fellow}$^{a b}$, L. Mankiewicz$^{a c}$

\vspace{1.5cm}

\begin{center}

{\em$^a$Physik Department, Technische Universit\"{a}t M\"{u}nchen, 
D-85747 Garching, Germany} 

{\em $^b$Petersburg Nuclear Physics Institute,
  188350, Gatchina, Russia
}

{\em $^c$N. Copernicus Astronomical Center, ul. Bartycka 18,
PL--00-716 Warsaw, Poland}

\end{center}

\vspace{1cm}

%{\em \today}

\vspace{3cm}

\centerline{\bf Abstract}
\begin{center}
\begin{minipage}{15cm}
  We have obtained a general solution of evolution equations for QCD
  twist-2 string operators in form of expansion over complete set of
  orthogonal eigenfunctions of evolution kernels in
  coordinate-space representation. In the leading logarithmic
  approximation the eigenfunctions can be determined using constraints
  imposed by conformal symmetry. Explicit formulae for the LO
  scale-dependence of quark and gluon twist-2 string operators are
  given.
\end{minipage}
\end{center}

%\vspace{1cm}
%{\em Submitted to Physics Letters B}
\end{center}
\end{titlepage}
\setcounter{footnote}{0}

\newpage

\noindent {\bf Introduction}\\

Recently, there has been a renewed interest in QCD evolution of skewed parton
distributions \cite{Ji97,Rad97}. Skewed parton distributions play a
crucial role in description of hard, exclusive QCD processes
\cite{Ji97,Rad97,CFS97,Ji98b,Blum99} which are actively considered
\cite{HERA,HERMES,COMPASS,TJNAF} as tools for investigation of new aspects of
non-perturbative QCD dynamics \cite{Ji98a,Guich98}. However, it is clear that
before a non-perturbative information can be reliably extracted from
experimental data, all perturbative aspects, such as QCD evolution, have to be
understood.  So far, the main effort has been devoted to studies of evolution equations for skewed parton distributions 
in the momentum representation
\cite{Ji97,Rad97,Blum99,FFGS97,Bel97b,MPW98a,Bel98a,Rad99,AMar99,Shuv99}.
This is certainly the most natural choice in the forward, deep inelastic
scattering limit, but, as observed recently \cite{KivMan99}, as soon as there
is a longitudinal momentum transfer between initial and final hadron states,
the corresponding amplitude can be as conveniently represented in terms of
momentum- as coordinate-space skewed parton distributions.

In a recent paper \cite{KivMan99} we have considered evolution of
charge-conjugation odd and even flavor non-singlet coordinate-space
skewed quark distributions in the leading-logarithmic (LO)
approximation \cite{BalBra88}. The resulting solutions have form of
a Neumann series expansions \cite{Erdeylyi} over a set of Bessel
functions of the first kind $J_{\frc{1}{2}+j}$, with $j$ assuming odd,
respectively even integer values.  That the Bessel functions enter the
game is not surprising at all -- it has been known for a long time
\cite{BalBra88} that they form a representation of the conformal group
in the coordinate-space and, as this symmetry is preserved at the LO
\cite{ER78,Conformal}, they simultaneously diagonalize corresponding
evolution kernels.  What is more interesting is that the resulting
expansions are valid uniformly for all values of the asymmetry
parameter, in contrast to the situation in the momentum-space. There, one has
to distinguish between cases where momentum fraction is smaller,
respectively larger than the asymmetry parameter. A compact formula,
based on the expansion in terms of local, multiplicatively
renormalizable operators, exists only in the former case
\cite{Rad97,Bel97b,Bel98a,Rad99,KivMan99}.

Evolution equations for twist-2 string operators have been known for a
long time \cite{BalBra88,Evolop}. Written in the coordinate-space
representation, they have been considered so far as an intermediate
step in derivation of evolution equations for skewed parton
distributions in the momentum space.  However, given the transparent
form of the solution to the problem of scale-dependence of
coordinate-space skewed parton distributions found in \cite{KivMan99}
it is natural to ask whether similar formulae can be obtained for
evolution of twist-2 string operators themselves. As we demonstrate
below, this problem has a positive answer. The main goal of the
present paper is to discuss an explicit solution to the evolution
equations for string operators in a form of expansion over orthogonal
eigenfunctions of corresponding evolution kernels.

As we consider expansion of a non-local operator in terms of an
orthogonal set of functions, it is natural that coefficients of such
an expansion are themselves non-local operators. Their scale
dependence is governed by a set of evolution equations which in
general result in a mixing between coefficients corresponding to
different orthogonal eigenfunctions. However, this term is present
only when the analysis goes beyond the leading order (LO)
approximation.

The remaining presentation is organized as follows. First, we
introduce a general solution to QCD evolution equations for twist-2
string operators in a form of expansion over a set of orthogonal
eigenfunctions of evolution kernels, find expansion coefficients and
discuss their scale-dependence.  Next, we perform
an explicit analysis of the evolution equations in the LO
approximation. The crucial point here is that appropriate sets of
orthogonal eigenfunctions can be determined using constraints imposed
by the conformal symmetry, .i.e. without solving the eigenvalue
problems. In this approximation, the expansion coefficients are
identical with conformal string operators introduced in
\cite{BalBra88}. As a result, we obtain explicit expressions for QCD
evolution of quark and gluon string
operators. Finally, we summarize.\\

\noindent {\bf Evolution equations for QCD string operators}\\
Let us start from a discussion of a general structure of QCD evolution
equations for string operators. As in Ref.\cite{BalBra88}, we introduce the
following definitions of quark, respectively gluon twist-2 operators
\be
\lab{eq:Q_op_def}
O(\alpha,\beta) = {\bar q}(\frac{\alpha+\beta}{2}z) \hat{z}
\left[\frac{\alpha+\beta}{2}z,\frac{\alpha-\beta}{2}z\right]
q(\frac{\alpha-\beta}{2}z)\, ,
\ee
\be
G(\alpha,\beta) =  z_{\alf} G_{\mu \alf}^{a}(\frac{\alpha+\beta}{2}z) 
\left[\frac{\alpha+\beta}{2}z,\frac{\alpha-\beta}{2}z\right]_{ab}
 G_{\mu \bt}^{b}(\frac{\alpha-\beta}{2}z) z_{\bt} \, .
\lab{gluon}
\ee 
In the above formula $z$ is a light-like vector, $z^2=0$.
The square brackets denote the 
path-ordered exponential:
\be
[az,bz] = \,
{\cal P} \exp [ -i g 
\-\int_a^b z_\mu A^{\mu}(tz) \, d t]
\ee
which ensures gauge-invariance of the above definitions.
Note that $\alpha\, z$ describes the center
of the string composed from quark fields and the gluon line between them while
$\beta\, z$ corresponds to its ``length'', understood simply as the
difference between coordinates of quark or gluon fields at the string ends.

Charge conjugation odd and even quark string operators are obtained from
 $O(\alpha,\beta)$ by taking its components symmetric, respectively
antisymmetric in $\beta$. Denoting the former and the latter by
$O^+(\alpha,\beta)$, respectively $O^-(\alpha,\beta)$ we define
\baa
\lab{eq:O_C_decomp}
O^+(\alpha,\beta) = \frac{1}{2}\left(O(\alpha,\beta) + O(\alpha,-\beta)
\right) \, ,
\nonumber \\
O^-(\alpha,\beta) = \frac{i}{2}\left(O(\alpha,\beta) - O(\alpha,-\beta)
\right) \, .
\eaa

Similarly to local operators, bare string operators have
UV-divergences and should be renormalized. In complete analogy with
the former case one can define renormalized string operator through
the relation:
\be
O_{R}(\om)=Z(\om,\om')\otimes O_{B}(\om') \, .
\lab{Ren} 
\ee
Here we have introduced a compact notation $\otimes$ for the two-dimensional integral
$$
\int_{-\infty}^{\infty} d\alpha \, \int_0^\infty d\beta
$$
and $\om\equiv \{\alf,\bt \}$. Function $Z(\om,\om')$ corresponds to a
renormalization constant in a local case. We suppose that in MS-scheme it has
the following structure:
\be
Z(\om,\om')=\delta(\om-\om')+\sum_{n\ge 1}\frac{1}{\eps^n}
\sum_{k\ge n} a_s^k Z_{nk}(\om,\om') \, ,
\ee
where $a_s = \frac{\alpha_s(\mu^2)}{4 \pi}$ and $\eps = 2 - d/2$, $d$ being
the space-time dimension. Coefficients $Z_{nk}$ can be
calculated 
in 
perturbation theory. In the MS-like schemes
they do not depend on the renormalization
scale $\mu$. 

The renormalization group equation can be derived by taking a total
derivative over $\mu$ of the left- and right-hand sides of
(\ref{Ren}). In this way one arrives at an evolution equation for the
renormalized string operator $O_R(\om)$ \cite{BalBra88,Evolop}
\be
\mu\frac{d}{d\mu}O_R(\om)=a_s V(\om,\om')\otimes O_R(\om') \, .
\lab{RGeq}
\ee     
The evolution kernel $V(\om,\om')$ is defined as
\be
\ba{l}
\displaystyle
a_s V(\om,\om')=\mu\frac{\mu}{d\mu}Z(\om,\om'')\otimes Z^{-1}(\om'',\om') \, .
\\[4mm]\displaystyle
Z^{-1}(\om'',\om')\otimes Z(\om',\om)=\delta(\om''-\om) \, .
\ea
\lab{kern}
\ee 
Although it has not been indicated
explicitly, $V(\om,\om')$ is a function of
the QCD coupling constant: 
\be
V(\om,\om')=\sum_{n\ge 0} \, a_s^n(\mu^2)V^{(n)}(\om,\om') \, ,
\lab{PTkern}
\ee 
except of the leading order (LO) approximation, when it is
given simply by
a function $V^{(0)}(\om,\om')$.

After these preliminaries, we shall construct now a general solution of the
operator evolution equation 
(\ref{RGeq}). To this end, let us assume that one can find a solution of the 
following system of equations for eigenfunctions and corresponding eigenvalues
of the evolution kernel:
\be
\ba{l} \displaystyle
V(\om,\om')\otimes\vphi_i(\om')=\gamma_i\vphi_i(\om')\\[2mm]
\displaystyle
\bar{\vphi}_i(\om)\otimes V(\om,\om')=\gamma_i\bar{\vphi}_i(\om)
\ea
\lab{eigen}
\ee
In the theory of integral equations the second equation is called 
conjugated to the 
first one \cite{IntEq}. 

Eigenfunctions
$\vphi_i(\om)$ and $\bar{\vphi}_j(\om)$ are orthogonal with respect to the
scalar product $\otimes$, i.e.
\be
\langle\bar{\vphi}_i,\ \vphi_j \rangle=
\bar{\vphi}_i(\om)\otimes\vphi_j=\delta_{ij}N_i\ ,
\lab{sprod}
\ee
where $N_i$ is some normalization constant. Note that beyond the leading order
eigenfunctions
$\vphi_i(\om)$, $\bar{\vphi}_j(\om)$ and eigenvalues $\gamma_i$ depend on the
running QCD coupling because the kernel $V(\om,\om')$ 
itself becomes a function
of $a_s(\mu^2)$, see eq.(\ref{PTkern}).
 
Solution of eq.(\ref{RGeq}) can be written in the following form:
\be
O_R(\om)=\sum_{n \ge 0} S_n\ \vphi_n(\om) \, .
\lab{sol}
\ee
The operator-valued coefficients $S_n$ can be obtained by taking the scalar
products of both sides of the above equation with eigenfunctions
$\bar{\vphi}_n$: 
\be
S_n=\frac{1}{N_n} \, \bar{\vphi}_n(\om') \otimes O_R(\om')
\lab{string} 
\ee
As it follows, $S_n$ themselves are non-local operators.
Moreover. eq.(\ref{RGeq}) leads to 
the following set of evolution equations for operators
$S_n$: 
\be
\mu\frac{d}{d\mu}S_n +
\sum_{m\ge 0}S_m\langle\bar{\vphi}_n ,\ \dot{\vphi}_m  \rangle
  =a_s\, \gamma_n S_n,\mskip10mu 
\dot{\vphi}_n\equiv \bt(\alf_s)\frac{\partial}{\partial\alf_s}\vphi_n \, .
\lab{Sevol}
\ee
The second term on the left-hand side induces mixing between operators
corresponding to different orthogonal eigenfunctions. It is absent in
the LO because eigenfunctions $\vphi_n$ do not depend on $\alf_s$ in
this approximation.  As a consequence operators $S_n$ become
multiplicatively renormalizable.
%This equations are equivalent to usual renormalization grope equations
%for local operators.
%At present we don't see arguments
%in support that this mixing should be triangular as it take place for the
%local operator.
This case will be discussed in more details in the next section.\\

\noindent{\bf LO evolution of string operators}\\
In the following we will follow notation of \cite{BalBra88}. To the
leading-order accuracy, the evolution kernel is given by
$V(\om,\om')=V^{(0)}(\om,\om')$ and is therefore independent on
$a_s(\mu)$. Arguments based on the conformal
symmetry can be used to find explicit solutions of the eigenvalue
equation (\ref{eigen}). In Ref.\cite{BalBra88} the nonlocal conformal
operators were introduced for the first time and the conformal
symmetry arguments were used to find the solution of (\ref{RGeq}) in
terms of an integral representation over a complex conformal spin $j$.
Our goal here is to represent a solution in the form of an expansion
(\ref{sol}). Is is equivalent to the former one, but much simpler from
the numerical point of view.

Now, let us consider solutions to the LO evolution equations
in the framework set up in the previous section.  First, note
that the two-dimensional eigenvalue problem (\ref{eigen}) can be
reduced to a one-dimensional one by using translational invariance to
separate the dependence on the variable $\alf$, which describes the
position of the center-of-mass of the string. This is done by
substituting
\be
\phi_j(\alf,\bt)=e^{ik\alf}f_j(k\bt),\
\bar{\phi}_j(\alf,\bt)=e^{ik\alf}\bar{f}_j(k\bt) \, .
\lab{decomp}
\ee 
In this way, $\phi_j(\alf,\bt)$
and $\bar{\phi}_j(\alf,\bt)$ can be interpreted as wave
functions of bound systems of two particles, with a momentum $k$ and an
internal motion described 
by $f_j(k\bt)$ and $\bar{f}_j(k\bt)$, respectively.
Let us first consider the case of flavor non-singlet quark operators.
Inserting (\ref{decomp}) into the eigenvalue equations (\ref{eigen}) one
arrives at the following system of equations \ci{BalBra88}:
\be
\ba{l} \displaystyle
H_1f_j \equiv \int_0^{\rho_2}V(\rho_1,\rho_2)f_j(\rho_1)=
\gamma_j f_j(\rho_2),
\\[4mm]\displaystyle
H_2\bar{f}_j \equiv \int_{\rho_1}^{\infty}V(\rho_1,\rho_2)\bar{f}_j(\rho_2)=
\gamma_j\bar{f}_j(\rho_1) \, .
\ea
\lab{Redeigen}
\ee
where $\rho\equiv k\beta$. The reduced kernel $V(\rho_1,\rho_2)$ is
given by 
\be 
V(\rho_1,\rho_2)= -3\delta(\rho_1-\rho_2) +
\frac{2\sin(\rho_1-\rho_2)}{\rho_2}+
\frac{4\rho_1\cos(\rho_1-\rho_2)}{\rho_2(\rho_1-\rho_2)}+
4\delta(\rho_1-\rho_2)\int_0^{\rho_2}\frac{d\rho}{\rho} \, .  
\ee 
Equations (\ref{Redeigen}) have the same form as the ERBL-equation for
evolution of the pion distribution amplitude \cite{ER78,BL79}. The latter is
defined as the reduced matrix element of twist-2 non-local quark operator
between the pion state and the vacuum. Conformal symmetry was shown to
play a crucial role in finding solution of the LO ERBL equation a long-time
ago \cite{ER78,BL79,Conformal}. In complete analogy, we shall now demonstrate
how the symmetry considerations allow to determine eigenfunctions $f_j(k\bt)$
and $\bar{f}_j(k\bt)$. Note that in the present case the evolution is
considered at the operator level, without any reference to matrix elements.

${\bf J}^2$, the Casimir operator of the collinear
conformal group SO(2,1), can be represented as an differential operator acting
on the parameters $\alf, \bt$.  After substitution (\ref{decomp}) one obtains
\cite{BalBra88}:
\be
\ba{l} \displaystyle
{\bf J}^2f_j(\rho)\equiv L_1f_j(\rho)=\left[
\rho^2\frac{d^2}{d\rho^2}+4\rho\frac{d}{d\rho}+\rho^2k^2+2               
\right]f_j(\rho)
\\[4mm] \displaystyle
{\bf J}^2\bar{f}_j(\rho)\equiv L_2\bar{f}_j(\rho)=\left[
\rho^2\frac{d^2}{d\rho^2}+\rho^2k^2\right]\bar{f}_j(\rho) \, .
\ea
\lab{J2}
\ee
Note that the explicit form of $J^2$ depends on whether it acts on
 $f_i(k\bt)$ or on $\bar{f}_i(k\bt)$. Using equations (\ref{Redeigen})
and (\ref{J2}) it is easy to check by an explicit calculation that
 ${\bf J}^2$ indeed commutes with the 'Hamiltonians' $H_{1,2}$, as
required by the symmetry arguments.  As it follows, the eigenfunctions of
 $H_{1,2}$ have to be the same as the eigenfunctions of $L_{1,2}$:
\be
 L_1f_j(\rho)=j(j+1)f_j(\rho),\ L_2\bar{f}_j(\rho)=j(j+1)\bar{f}_j(\rho) \, ,
\lab{J2eigen}
\ee
where $j$ are positive integers.
As $L_1$ and $L_2$ are related by
\be
L_2\rho^2=\rho^2L_1
\lab{rel}
\ee
one obtains the following relation between $f_j(\rho)$ and
 $\bar{f}_j(\rho)$ 
\be
\bar{f}_j(\rho) = \rho^2 f_j(\rho) \, .
\ee

Solutions of the set of differential equations (\ref{J2eigen}) are
given by Bessel functions:
\be
f_j(\rho)=c_j(k)\rho^{-3/2}Z_{j+1/2}(\rho), 
\bar{f}_j(\rho)=\bar{c}_j(k)\rho^{1/2}Z_{j+1/2}(\rho) \, ,
\lab{Bessel}
\ee
where $c_j(k),\ \bar{c}_j$ are yet undetermined normalization
constants.  As the operator $O_R(\omega)$ has a well-defined local
limit as $\beta \to 0$, the validity of expansion (\ref{sol}) requires
that $f_j(\beta)$ are regular in this limit as well. In terms of $f_j$
and $\bar f_j$ the scalar product (\ref{sprod}) reduces to
\be
\langle\bar{f}_j(\rho)f_i(\rho)  \rangle=
\int_0^{\infty}d\rho\bar{f}_j(\rho)f_j(\rho)=N_i\delta_{ij} \, ,
\lab{ort}
\ee
and therefore $f_j$
and $\bar f_j$ must be sufficiently regular such that the integral exists.
These conditions are satisfied when we identify
$Z_{j+1/2}(\rho)$ with $J_{j+1/2}(\rho)$, the Bessel functions of the
first kind. In particular, the scalar product (\ref{ort}) reduces to
the well known integral ($\{i,\ j\}=1,2,3,...$) \cite{Ryzhik}:
\be
\langle\bar{f}_i(\rho)f_j(\rho)  \rangle=\bar{c}(k)_ic_j(k)
\int_0^{\infty}\frac{d\rho}{\rho}J_{j+1/2}(\rho)J_{i+1/2}(\rho)=
\bar{c}(k)_i c_j(k)\frac{\delta_{ij}}{2i+1} \, .
\lab{int}
\ee  
For reasons which will shortly become clear we set
 $\bar{c}_j(k)=k^{-1/2}$ and $c_j(k)=k^{3/2}$.  With this choice the
nonlocal operators $S_j$ defined according to (\ref{string}) assume
the form:
\be\label{eq:conf_q_op}
S_j \equiv S(\frc{1}{2}+j,k;\mu^2) = 
\int_{-\infty}^{\infty} d\alpha \, e^{i k \alpha}\,
\int_0^\infty d\beta \, \sqrt{\beta} J_{\frc{1}{2}+j} (|k|\beta)
\, O(\alpha,\beta)_{\mu^2} \, .
\ee
As 
noted above, they are multiplicatively renormalizable \cite{BalBra88}:
\be
\ba{l}\displaystyle
\mu\frac{d}{d\mu}S(\frc{1}{2}+j,k;\mu^2) = 
a_s\gamma_j S(\frc{1}{2}+j,k;\mu^2),\,
S(\frc{1}{2}+j,k;\mu^2)=L_{j}S(\frc{1}{2}+j,k;\mu^2_0),\\[4mm]
L_{j}\equiv 
\left(\frac{\log{(\mu^2/\Lambda^2)}}{\log{(\mu^2_0/\Lambda^2)}}\right)
^{-\frac{\gamma(j)}{b_0}},\, 
\gamma(j)\equiv\gamma_{QQ} = C_F\,\left(3 + \frac{2}{j(j+1)} -
 4 (\Psi(j+1) + \gamma_{\rm E})\right) \, .
\ea
\lab{Mult}
\ee
Here, $\gamma(j)$ is the corresponding anomalous dimension known, e.g., from
analysis of deep-inelastic scattering process.

Now we are in position to write down the decomposition (\ref{sol})
explicitly:  
\be
O(\alf,\bt)_{\mu^2}=\int_{-\infty}^\infty \frac{dk}{2\pi}\, e^{-ik\alpha}
{\beta^{-\frac{3}{2}}} \sum_{j=1}^\infty (1+2j) L_{j} 
J_{\frc{1}{2}+j}(|k|\beta) S(\frc{1}{2}+j,k;\mu^2_0)
\lab{Jexp}
\ee   
It remains only to check the normalization. As 
$\mu^2=\mu^2_0$ and therefore 
$L_{j}=1$ for all $j$ we should have
\be
O(\alf,\bt)=\int_{-\infty}^\infty d\alf'\int_0^\infty d\bt'
{\cal R}(\alf,\bt|\alf',\bt')\, O(\alf',\bt')
\ee
with
\be
{\cal R}(\alf,\bt|\alf',\bt')=\int \frac{dk}{2\pi}e^{ik(\alf'-\alf)}
{\beta^{-\frac{3}{2}}}{\beta'^{\frac{1}{2}}}
\sum_{j=1}^\infty (1+2j)J_{\frc{1}{2}+j}(|k|\beta)J_{\frc{1}{2}+j}(|k|\beta')=
\delta(\alf-\alf')\delta(\bt-\bt') \, .
\lab{resolv}
\ee
To this end, recall that the series
\be
{\beta^{-\frac{3}{2}}}{\beta'^{\frac{1}{2}}}
\sum_{j=1}^\infty (1+2j)J_{\frc{1}{2}+j}(|k|\beta)J_{\frc{1}{2}+j}(|k|\beta')=
\delta(\bt-\bt')
\ee
does not depend on $k$ and represents the Neumann expansion of a
$\delta$-function in terms of Bessel functions
\cite{KivMan99,Erdeylyi}. Equation (\ref{resolv}) follows immediately.
Note that this observation explains also the choice of the
coefficients $c_i ,\, \bar{c_i}$.

Equation (\ref{Jexp}) describes solution of the LO evolution equation
as an expansion in terms of orthogonal eigenfunctions of the evolution
kernel. On the other hand, it can
be interpreted as an expansion of a quark string operator in terms of
conformal string operators. Series expansion in terms of Bessel
functions is known in mathematical literature as the Neumann series,
see \cite{Erdeylyi} for more detailed information. Evolution of
flavor non-singlet, charge-conjugation odd and even quark string
operators can be obtained from (\ref{Jexp}) by taking its symmetric,
respectively antisymmetric in $\beta$ components.

Conformal operators (\ref{eq:conf_q_op}) are defined only for
positive, integer values of the conformal spin $j$. On the other hand, the
solution found in 
\cite{BalBra88} is based on an integral
representation which requires analytical continuation of $j$
into the complex plane. In our previous paper \cite{KivMan99} we have
explicitly checked that integration 
over $j$ indeed reproduces the series  (\ref{Jexp}).

Let us now turn our attention to operators with flavor singlet and
positive charge parity quantum numbers. Here we have to consider
mixing between gluon and quark string operators, see equations
(\ref{gluon}) and (\ref{eq:O_C_decomp}) for corresponding definitions.
Note that the quark operator $Q \equiv O^-$ receives contribution from all
$N_F$ active flavors.  An analysis analogous to the flavor
non-singlet case leads to the following basis of the nonlocal
conformal quark and gluon operators:
\be
\ba{l}\displaystyle
S_G(1/2+j,k;\mu^2)=\int_{-\infty}^{\infty} d\alpha \, e^{i k \alpha}\,
\int_0^\infty d\beta \, \beta^{3/2} J_{\frc{1}{2}+j} (|k|\beta)
\, G(\alpha,\beta)_{\mu^2} \, ,
\\[4mm]\displaystyle
S_Q(1/2+j,k;\mu^2)=\int_{-\infty}^{\infty} d\alpha \, e^{i k \alpha}\,
\int_0^\infty d\beta \, \sqrt{\beta} J_{\frc{1}{2}+j} (|k|\beta)
\, Q(\alpha,\beta)_{\mu^2} \, .
\ea
\lab{Cbasis}
\ee
Here $j=2+2n,\, n=0,1,2,3,...$ is the conformal
spin. Operators $S_G(1/2+j,k;\mu^2)$ and
$S_Q(1/2+j,k;\mu^2)$ depend on the renormalization scale according to
\be
\ba{l}\displaystyle
S_G(1/2+j,k;\mu^2)=\left[\lambda_{+}L_{+}-\lambda_{-}L_{-} \right]
S_G(1/2+j,k;\mu^2_0)+
\\[4mm]\displaystyle \mskip80mu 
+\gamma_{GQ}\frac{(j-1)}{\sqrt{D}}\left[L_{+}-L_{-} \right]
S_Q(1/2+j,k;\mu^2_0) \, ,
\ea
\lab{Egluon}
\ee
\be
\ba{l}\displaystyle
S_Q(1/2+j,k;\mu^2)=\left[\lambda_{+}L_{-}-\lambda_{+}L_{-} \right]
S_Q(1/2+j,k;\mu^2_0)+
 \\[4mm]\displaystyle \mskip80mu
+\frac{\gamma_{QG}}{(j-1)}\, \frac{1}{\sqrt{D}}\left[L_{+}-L_{-} \right]
S_G(1/2+j,k;\mu^2_0) \, .
\ea
\lab{Equark}
\ee
Here we have introduced a shorthand notation:
\be
\ba{l}\displaystyle
L_{\pm}=\left(\frac{\log{(\mu^2/\Lambda^2)}}{\log{(\mu^2_0/\Lambda^2)}}\right)
^{-\frac{\gamma_{\pm}(j)}{b_0}},\,
\lambda_{\pm}=\frac12\frac1{\sqrt{D}}[\gamma_{GG}-\gamma_{QQ}\pm\sqrt{D}],
\\[6mm]\displaystyle
\gamma_{\pm}=\frac12\left(\gamma_{GG}+\gamma_{QQ}\pm\sqrt{D}\right),\
 D=(\gamma_{GG}-\gamma_{QQ})^2+4\gamma_{GQ}\gamma_{QG} \, .
\ea
\ee 
In convention adopted in this paper the anomalous dimensions
$\gamma_{GG},\gamma_{GQ},\gamma_{QG}$ differ by a sign from those
given in Ref.\cite{BalBra88}. Note that although not explicitly
indicated, $j$-dependence of the anomalous dimensions is implied
here.

To find the evolution of the string operators $G(\alf,\bt),\,
Q(\alf,\bt)$ we have to rewrite them as series expansions in terms
of the conformal string operators (\ref{Cbasis}). Consider first the
gluon operator $G(\alpha,\beta)$. Using the Neumann expansion of the
$\delta$-function:
\be
\bt\delta(\bt-\bt')= \sum_{n=0}^\infty (2\nu+2+4n) 
J_{\nu+1+2n}(|k|\beta)J_{\nu+1+2n}(|k|\beta^\prime) \, ,
\ee
one easily obtains the desired formula:    
\be
\ba{l}\displaystyle
G(\alpha,\beta;\mu^2)= \int_{-\infty}^\infty d\alpha^\prime  \int_0^\infty
d\beta^\prime\, \delta(\alpha-\alpha^\prime) \delta(\beta-\bt')
G(\alpha',\beta';\mu^2)=
\\[4mm]\displaystyle
=\int\mskip-3mu d\alpha^\prime\mskip-3mu  \int\mskip-3mud\beta^\prime
\int\mskip-3mu \frac{dk}{2\pi}e^{ik(\alpha'-\alpha)}
\bt^{-\frac52}\sum_{n=0}^\infty (5+4n) 
J_{\frac52+2n}(|k|\beta)J_{\frac52+2n}(|k|\beta')\bt^{\prime\frac32}
G(\alpha',\beta';\mu^2)=
\\[4mm]\displaystyle
=\int_{-\infty}^\infty \frac{dk}{2\pi}e^{-ik\alpha}\bt^{-5/2}\sum_{n=0}^\infty
(5+4n) J_{\frac52+2n}(|k|\beta)S_G({\textstyle\frac52}+2n,k;\mu^2) \, .
\ea
\lab{gl_con_exp}
\ee
Taking into account (\ref{Egluon}) one finally arrives at the
following expression for the evolution of the gluon operator:
\be
\ba{l}\displaystyle
G(\alpha,\beta;\mu^2)=
\int_{-\infty}^\infty \frac{dk}{2\pi}e^{-ik\alpha}\bt^{-5/2}\sum_{n=0}^\infty
(5+4n) J_{\frac52+2n}(|k|\beta)\times
\\[4mm]\displaystyle \mskip80mu 
\times\biggl\{
\left[\lambda_{+}L_{+}-\lambda_{-}L_{-} \right]_{(j=2n+2)}
S_G({\textstyle\frac52}+2n,k;\mu^2_0)+
 \\[4mm]\displaystyle \mskip80mu 
+\gamma_{GQ}\frac{2n+1}{\sqrt{D}}\left[L_{+}-L_{-} \right]_{(j=2n+2)}
S_Q({\textstyle\frac52}+2n,k;\mu^2_0)\biggr\} \, .
\lab{gluon_evol}
\ea 
\ee

Evolution of the quark-singlet operator can be found in a similar way.
First, we establish an expansion in terms of non-local conformal
operators
\be
Q(\alpha,\beta;\mu^2)=\int_{-\infty}^\infty \frac{dk}{2\pi}e^{-ik\alpha}
\bt^{-3/2}\sum_{n=0}^\infty(5+4n) J_{\frac52+2n}(|k|\beta)
S_Q({\textstyle\frac52}+2n,k;\mu^2) \, .
\lab{Q_conf_exp}
\ee
The resulting solution to the evolution equations reads
\be
\ba{l}\displaystyle
Q(\alpha,\beta;\mu^2)=\int_{-\infty}^\infty \frac{dk}{2\pi}e^{-ik\alpha}
\bt^{-3/2}\sum_{n=0}^\infty(5+4n) J_{\frac52+2n}(|k|\beta)\times
\\[4mm]\displaystyle \mskip80mu
\times\biggl\{
\left[\lambda_{+}L_{-}-\lambda_{-}L_{+} \right]_{(j=2n+2)}
S_Q({\textstyle\frac52}+2n,k;\mu^2_0)+
 \\[4mm]\displaystyle \mskip80mu
+\frac{\gamma_{QG}}{(2n+1)}\, \frac{1}{\sqrt{D}}\left[L_{+}-L_{-}
\right]_{(j=2n+2)} 
S_G({\textstyle \frac52}+2n,k;\mu^2_0)\biggr\} \, .
\lab{quark_evol}
\ea 
\ee 

Equations (\ref{Jexp}), (\ref{gluon_evol}) and (\ref{quark_evol}) are new and
represent the main results of this paper. As advocated in the previous
section, they describe the scale dependence of QCD string operators in form of
expansion over a set of eigenfunctions of corresponding evolution kernels.
Evaluating matrix elements between appropriate nucleon states one finds the
explicit form of scale dependence of skewed parton distributions in the
coordinate-space representation \cite{KivMan99}. Due to translational
invariance, the integral over $k$ can be trivially performed and one is left
with an expansion of corresponding coordinate-space skewed parton
distributions in terms of orthogonal set of Bessel functions, valid for all
values of the asymmetry parameter. Hence, the solution in the coordinate-space
is much more transparent than in the
momentum representation.\\

\noindent {\bf Summary}\\
We have argued that solutions of evolution equations for twist-2 QCD string
operators can be written as series
expansions in terms of the orthogonal sets of eigenfunctions of evolution
kernels. We have obtained a general form of evolution equations for
operator-valued expansion coefficients and found that in a general case
evolution results in a mixing between coefficients corresponding to different
eigenfunctions.
 
In the LO approximation the eigenfunctions of evolution kernels can be found
explicitly using conformal symmetry arguments and, at the same time,
evolution equations for expansion coefficients become much simpler than in the
most general case. As a result the expansion coefficients can be identified
with conformal string operators.  Those with flavor non-singlet or
charge-parity odd quantum numbers are simply multiplicatively renormalized and
those corresponding to flavor singlet and positive charge-parity are subject
only to the usual mixing between gluon and quark operators.  In both cases
explicit formulae for scale dependence of quark and gluon string operators
have been derived.

Taking corresponding matrix elements one obtains immediately solution to the
problem of scale dependence of skewed parton distributions in the
coordinate-space. Here, contrary to the situation in momentum space, the
resulting expansion over a set of orthogonal eigenfunction is valid for all
values of the asymmetry parameter.

\vspace{1cm}

%\newpage
        
\noindent
{\bf Acknowledgments:} 

\noindent
N.K. would like to thank the Humboldt Foundation for their financial
support. We gratefully acknowledge discussions with V. Braun, G. Piller,
and A. Radyushkin. 
\\

\end{document}